\documentclass[twocolumn,aps,prd,longbibliography,groupedaddress,nofootinbib]{revtex4}

\input{config}

\begin{document}

\title{Relational structures of fundamental theories}

\author{Pierre Martin-Dussaud}
\email{}
\affiliation{Institute for Gravitation and the Cosmos, The Pennsylvania State University, University Park, Pennsylvania 16802, USA}
\affiliation{Basic Research Community for Physics e.V., Mariannenstra\ss e 89, Leipzig, Germany}

\date{ \small\today}

\begin{abstract}
\noindent
General relativity and quantum mechanics have both revealed the relativity of certain notions that were previously thought to be absolute. I clarify the precise sense in which these theories are relational, and I argue that the various aspects of relationality pertain to the same movement in the progress of physical theories.
\end{abstract}

\maketitle 

It is a common place to point at the many differences between general relativity (GR) and quantum mechanics (QM). The gap between these theories motivates the construction of a bridge that one would call \textit{quantum gravity}. To achieve such a quest it is valuable to look, not only at the differences, but also at the points where the two theories can meet. For instance, physicists study black holes or the early cosmology, where both realms of physics come into play. Here I would like to ask the question at a more conceptual level, and focus on a notion which appears in both theories, although it takes different aspects: the notion of \textit{relationality}.

In theories of space-time, the word "relativity" is used more often than "relationality", but this is a matter of habit, rather than of clear difference of meaning in common language. I will use both terms interchangeably, with a slight preference for relationality, because it is less connoted --in fact, still a neologism for most dictionaries. Relationality appears crucially in GR, where it takes the form of diffeomorphism invariance. It also appears in an essential manner in QM, at least if one is ready to buy a relational interpretation of it, a stance to which I stick to. One may wonder whether the two notions of relationality are actually the same. Or to put it maybe more precisely, which kind of relationality should we expect at the level of quantum gravity? This question, first raised by Rovelli in \cite{rovelli1999}, was the starting point the following work.

In this paper, I do not reach many definite answers, but I try to be conceptually clear and to ask relevant questions. The main goal is to reach precision about what is meant by relationality, because it is a very general notion, and it is easy to get confused by a misuse of terms, especially when physicists from different communities use the same word with different meanings. I proceed gradually in four sections:
\begin{enumerate}
    \item [\ref{sec:classical_mechanics}] kick-off analysis with classical mechanics,
    \item [\ref{sec:general_relativity}] the novel insight of general relativity,
    \item [\ref{sec:relational_quantum_mechanics}] the role of the observer in quantum mechanics,
    \item [\ref{sec:conclusion}] take-away for epistemology.
\end{enumerate}

\section{Classical mechanics}
\label{sec:classical_mechanics}

Let’s start with a familiar example: the principle of relativity in classical mechanics. In his book \textit{Questions on the Four Books on the Heavens and the World of Aristotle}, the 14th century philosopher Buridan writes
\begin{quote}
\textit{If anyone is moved in a ship and he imagines that he is at rest, then, should he see another ship which is truly at rest, it will appear to him that the other ship is moved. This is so because his eye would be completely in the same relationship to the other ship regardless of whether his own ship is at rest and the other moved, or the contrary situation prevailed.} (\cite{barbour2001}, p. 203)
\end{quote}
It is surprising how closely Buridan's ideas are from the formulation of galilean relativity, as it is taught at school. Did people know about galilean relativity before Galileo himself? Not really. What Buridan is describing is \textit{kinematic relativity}. It is the observation that whenever one tries to describe concretely a motion, it has to be done with respect to some reference, which is postulated at rest. And this is true whatever the kind of motion, not only uniform straight line motion. Kinematic relativity is almost a linguistic fact about what motion means, and it was observed much before Galileo.

This initial observation raises an issue, that Julian Barbour, in \cite{barbour2001}, calls the \textit{fundamental problem of motion}: if all motion is relative and everything in the universe is in motion, how can one ever set up a determinate theory of motion?

\begin{figure}[h]
\centering
\includegraphics[width = 1 \columnwidth]{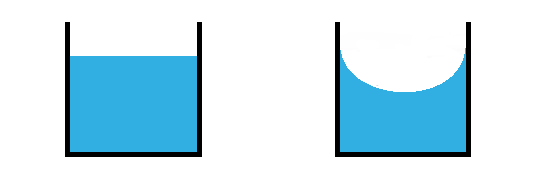}
\caption{Newton's bucket.}
\label{fig:bucket}
\end{figure}

The answer is provided by the thought experiment of Newton’s bucket. Imagine a bucket of water, hanged at a gallows and suppose that the water is not moving with respect to the bucket. This can happen in two cases (see figure \ref{fig:bucket}):
\begin{enumerate}
\item The bucket is hanging still, and the surface of the water is flat.
\item The bucket is turning with respect the gallows, and the surface of the water is concave.
\end{enumerate}
In absence of relative motion between the water and the bucket, observing the shape of the surface of the water, whether it is flat or concave, indicates whether the bucket is turning or not. One could say that this rotation is motion with respect to the gallows. But forget about the gallows and imagine that there is nothing else in the universe other than the bucket of water. Without any relative motion between the water and the bucket, the sole geometry of the surface tells whether the bucket is turning or not. With respect to what? It looks like there is no need of a reference system to say that the bucket is turning. But for Newton it must be turning with respect to something, and he concludes by calling this something \textit{absolute space}. Motion with respect to absolute space is called \textit{absolute motion}. Such a statement, which could be discarded as metaphysical, is actually essential, as it makes Newton's laws meaningful. Indeed, the equation of dynamics are written with coordinates which carry the physical meaning of a cartesian mapping of absolute space. This solves the fundamental problem of motion. It is remarkable that the introduction of the notion of absolute space is motivated by the faith in kinematic relativity: the bucket shall turn with respect to \textit{something}.

Now comes galilean relativity. I deliberately reverse chronological order to restore conceptual primacy. Galilean relativity states a fundamental experimental restriction on the description of absolute motion: although one can indeed detect absolute rotation, as the example of the bucket shows, it is impossible to distinguish absolute rest from absolute uniform straight line motion. So galilean relativity is not the kinematic relativity of Buridan! Kinematic relativity is really a kinematical statement, i.e. it is about how motion can be described in general, while galilean relativity is a dynamical statement, as it is about how motion can be predicted: it claims that the dynamical laws take the same form within the class of uniform straight line motions. Galilean relativity does not say that absolute uniform straight line motion does not exist, but it says that it is undetectable, which is not strictly the same thing. 

To put it in a nutshell, Newton’s theory of motion is a sandwich of absolutism between two layers of relativism.
\begin{enumerate}
\item \textit{Kinematic relativity}: motion is motion with respect to something.
\item \textit{Absolute space}: there is a preferred reference system, space itself, and motion with respect to it is called absolute.
\item \textit{Galilean relativity}: absolute uniform straight line motion is undetectable.
\end{enumerate}
The layer of "absolute space" is crucial to distinguish kinematic relativity from galilean relativity.

This warming-up has shown that the relational aspects of a theory that we believe to know well are already delicate to grasp. We can now move on to more modern theories.

\section{General Relativity}
\label{sec:general_relativity}

\begin{figure}[h]
\centering
\includegraphics[width = 1 \columnwidth]{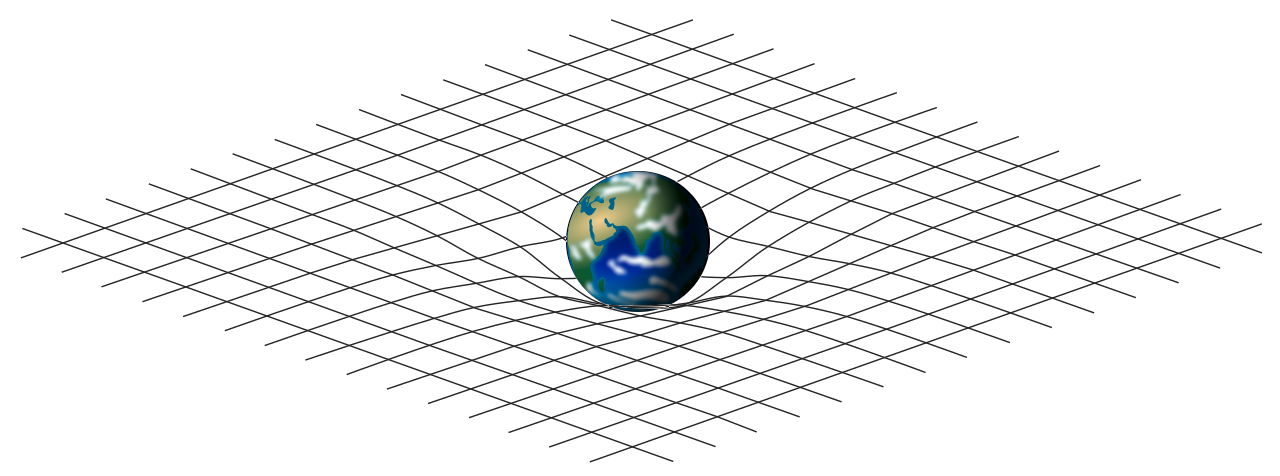}
\caption{A manifold curved by masses.}
\label{fig:curved-space-time}
\end{figure}

In GR, the common view in many textbooks is that space-time is a differentiable manifold over which a metric is defined. The picture that one has in mind is that of a grid, curved by masses (see figure \ref{fig:curved-space-time}).
The problem with this image is that it lets people think that space-time \textit{is} the continuum of points of the underlying topological manifold. These points would be basic entities of the world, on top of which would come an additional metric structure to measure the distance between the points. This view is false, because it overlooks a central aspect of GR, which is its relational novelty with respect to classical mechanics.

The three layers structure of relationality in classical mechanics might be transposable to GR. The first layer (kinematic relativity) is still there: it is the possibility to choose any grid to attribute numbers to points. It's called \textit{general covariance}, induced by \textit{passive} diffeomorphisms. The third layer (galilean relativity) is extended to accelerated motion (from a ship to a lift), and is then called \textit{diffeomorphism invariance}, induced by \textit{active} diffeomorphisms. General covariance and diffeomorphism invariance are conceptually different, but formally the same, which may convey the impression that there is not much room for any notion of absolute spacetime in between.

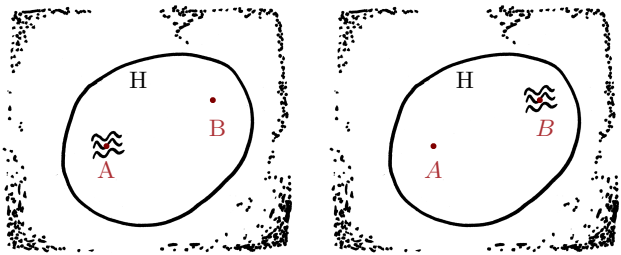
\begin{figure}[h]
\begin{overpic}[width = 0.49 \columnwidth]{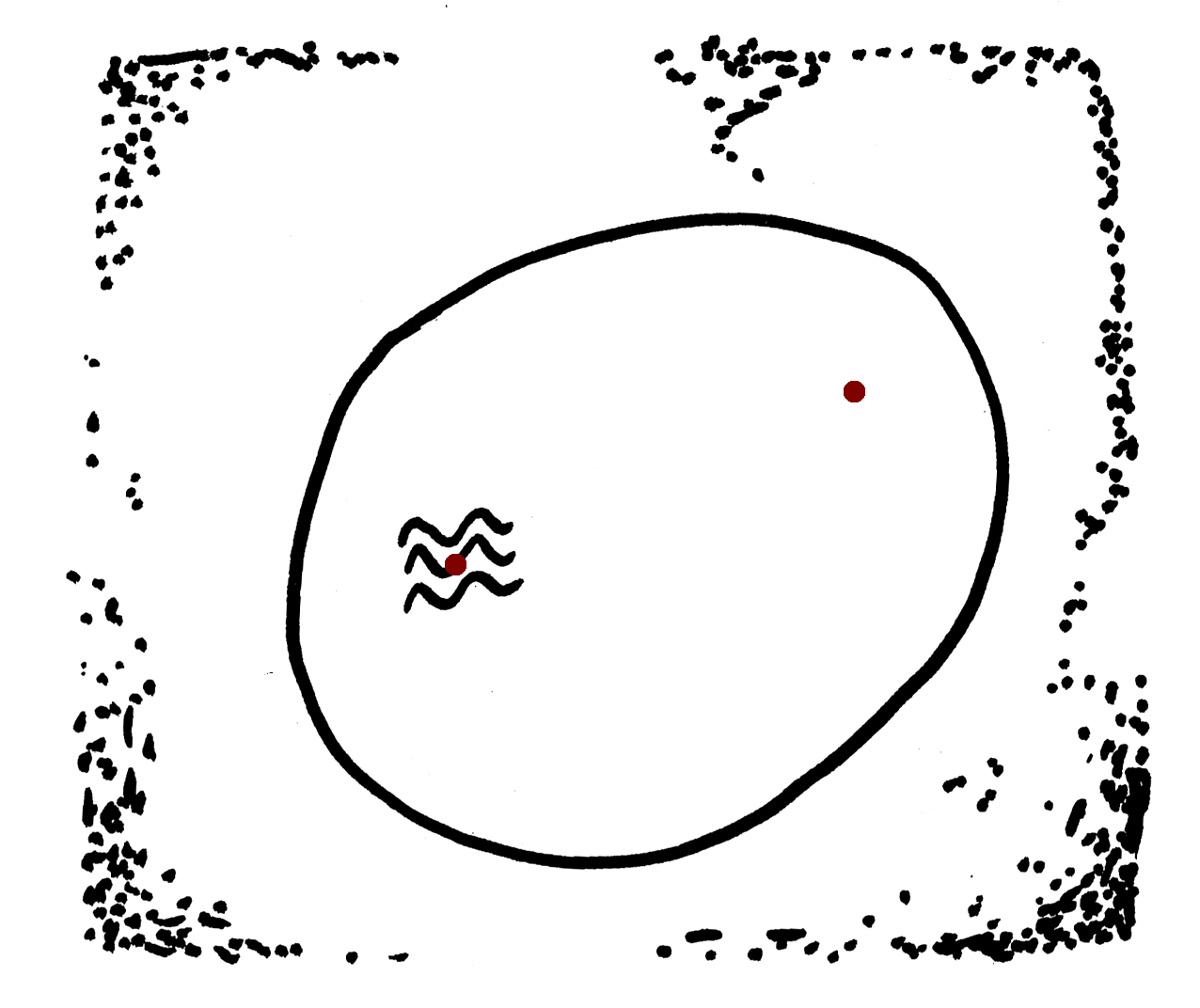}
\put (35,27) {\color{Maroon} A}
\put (70,40) {\color{Maroon} B}
\put (45,55) {H}
\end{overpic}
\begin{overpic}[width = 0.49 \columnwidth]{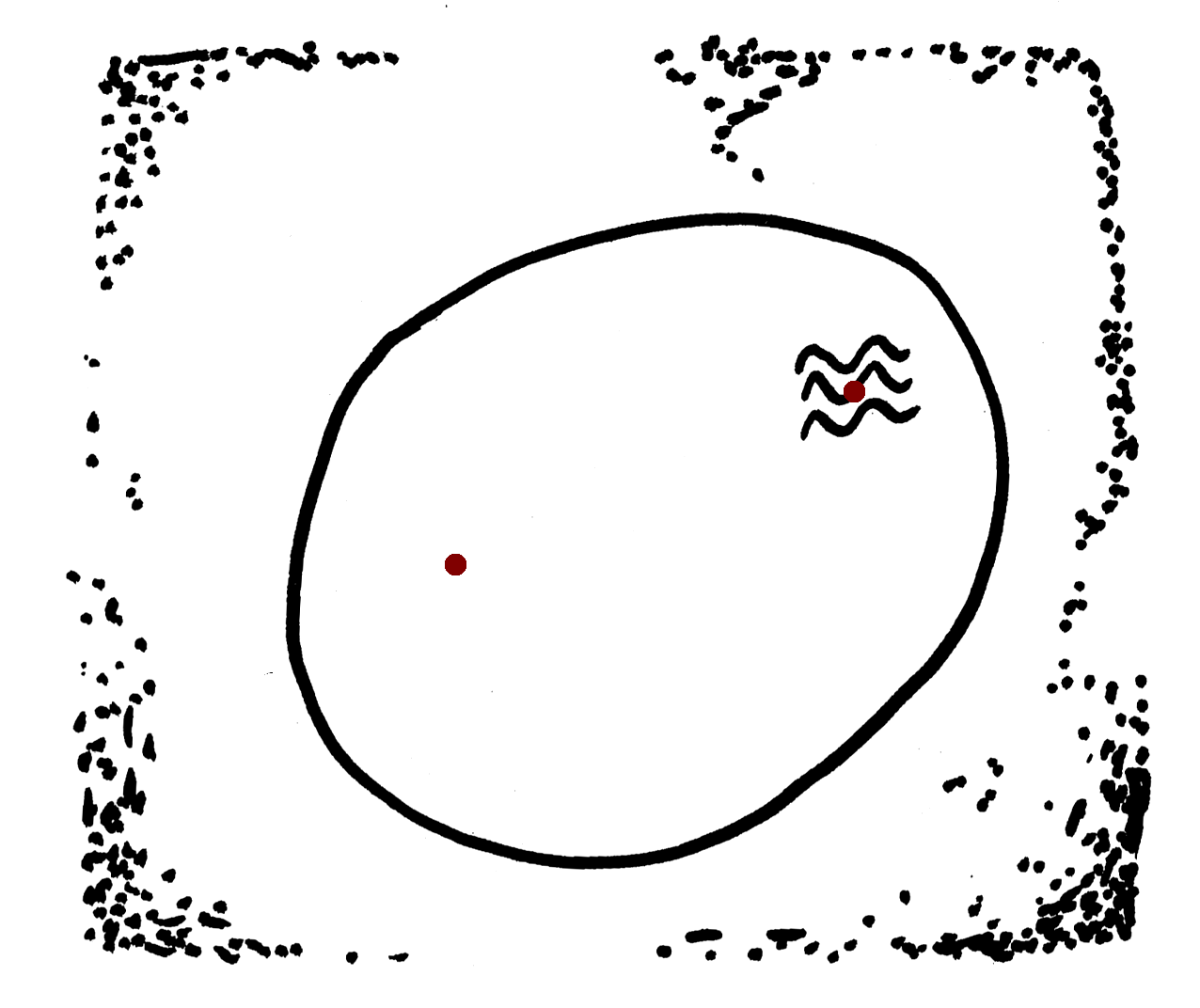}
\put (35,27) {\color{Maroon} $A$}
\put (70,40) {\color{Maroon} $B$}
\put (45,55) {H}
\end{overpic}
\caption{The hole argument. Left: space-time with metric $g_{\mu \nu}$. Right: space-time with metric $\tilde{g}_{\mu \nu}$.}
\label{fig:hole}
\end{figure}

Indeed, there is no possible physical existence to be attributed to the points of the space-time manifold. This is the consequence of the so-called \textit{hole argument}. Imagine some space-time with a given distribution of matter and a compact empty region $H$ (the hole), like figure \ref{fig:hole}.
Consider two points $A$ and $B$ in $H$. Although $H$ is empty, a metric $g_{\mu \nu}$, solution of Einstein's equation, may describe a curved geometry in $A$ (a gravitational wave), and a flat one in $B$, as in the left figure. There exist active diffeomorphisms, which leave the metric invariant outside the hole, but give a new metric $\tilde{g}_{\mu \nu}$ inside, which is flat in $A$ and curved in $B$, as in the right figure. According to the diffeomorphism invariance of the theory, $g_{\mu \nu}$ and $\tilde{g}_{\mu \nu}$ describe both the same physical situation, so that one must conclude that the statement "space-time is curved in $A$" is not physically meaningful. The manifold picture, made of points like $A$, is an absolute representation of space-time, in the sense that it conveys the impression that there is a preferred substratum with respect to which I can define a notion of absolute motion, like in Newton’s theory. The hole argument shows that this view is not viable.

The conclusion is hard to admit, because it breaks the widespread metaphysical postulate of realism. Realism is the stance that there is a one-to-one correspondence between the mathematical entities of a theory and the physically existing objects. The relational aspect of GR comes to challenge this view. The breakdown of realism does not mean that it is impossible to describe physics with maths, but it makes the correspondence indirect: it is harder to say what is observable and what is not. In Newton's theory, the coordinates $\vec{x}$ have an operational meaning: it is implicitly recognised that there is a concrete procedure to measure their values. On the contrary, in GR, the coordinates $x^\mu$ do not carry such an interpretation. Instead, the observables of GR are independent of the coordinates $x^\mu$.

So the manifold cannot be seen as an absolute space-time like in Newton’s theory. However, it is hard to say precisely the positive content of this fact: if space-time is not a continuum of points, what is space-time?

An easy answer --but false-- would be to say that there is no space-time, there are only distances between material points. This is false because Einstein equations admit vacuum solutions, which are non-trivial metrics describing an empty space-time. You can have gravitational waves for instance, which may carry enough energy to make you fall.

There is a way to catch the remaining absolute core of GR, by saying that physical space-time is an equivalence class of solutions of Einstein equations, two representatives being related by a diffeomorphism. This view challenges our mental representations, far from the naive image of figure \ref{fig:curved-space-time}, because the familiar notion of locality gets drowned into the abstract and global notion of equivalence class. In a similar spirit, the observables of general relativity, defined as diffeomorphism-invariant functions of the metric, appear to be very non-local quantities, like Wilson loops.

Another way to grab the physical content of GR, without wiping out locality, is to introduce some other field of reference on the manifold. This is the standard way to attribute meaning to the coordinates: use four scalar matter fields, or four satellites \cite{rovelli2002b}. In this context, space-time is deeply relational, because the values taken by a metric field $g_{\mu \nu}$ are only meaningful with respect to some other (matter) field. For instance, the Ricci scalar $R(x)$ is not an observable per se, unless $x$ is attributed a physically grounded meaning, like the place where some matter field $\phi$ takes some value. The role of the underlying topological manifold is only to provide a common ground where to compare various fields "at the same point". Thus, matter fields enable to recover the dubious notion of "absolute space". Newton's bucket can then be understood as motion with respect to the local gravitational field.

The two ways of extracting the physical content of GR, either by building diffeomorphism invariant quantities, or by providing a physical meaning to the coordinates with the help of an external field, arise from the same matter of fact: the numerical values of $g_{\mu \nu}(x)$ do not mean anything alone. The physics is extracted by comparing these values either to matter fields at the same point $x$, or to the same field $g_{\mu \nu}$ but at other points of the manifold (like along a loop). This is the relational lesson of GR.

\section{Relational Quantum Mechanics}
\label{sec:relational_quantum_mechanics}

Now let’s turn to QM. The fact that QM exhibits relational aspects is not universally acknowledged. One reason may be that it is not explicit in the name, contrary to "general relativity". What is explicit in the name of "quantum mechanics" is the idea of discreteness, which was indeed its starting point. The relational aspect of it was only recognised later.

\begin{figure}[h]
\centering
\begin{overpic}[width = 0.8 \columnwidth]{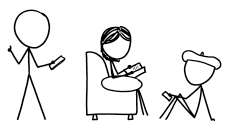}
\put (12,5) {A}
\put (45,10) {B}
\put (93,15) {C}
\end{overpic}
\caption{Wigner's friends. Image from \cite{xkcd}.}
\label{fig:gaston}
\end{figure}

It starts in 1957 with an article by Hugh Everett entitled \textit{‘Relative State’ formulation of Quantum Mechanics} \cite{everett1957}. He remarks the following fact that has been much discussed later under the name of Wigner's friend\footnote{The original article of Wigner dates from 1961 \cite{wigner1961}.}: consider a first observer B that carries on a measurement over a system C, and consider a second observer A that describes the overall situation without interacting with it (see figure \ref{fig:gaston}).
Denote $\rho^0_C$ the initial state of C (written as a density matrix). The two observers A and B describe a different evolution for C:
\begin{enumerate}
    \item [A.] The overall evolution is unitary and the state of C becomes
\begin{equation}
\rho^1_C = \Tr_B \left[ U (\rho^0_B \otimes \rho^0_C ) U^{-1} \right]
\end{equation}
with $U$ a unitary operator and $\rho^0_B$ the initial state of $B$.
    \item [B.] The state of C collapses to 
\begin{equation}
\rho^1_C = P \rho^0_C P^{-1},
\end{equation}
with $P$ a projector on one of the eigenstates of the observable.
\end{enumerate}
The two final states are different.

This thought experiment has given rise to a large amount of literature. It has notably fuelled many-world interpretations. According to them, at the moment of the measurement, the world splits in two branches, so that observer A also splits in two, each branch corresponding to a different result. Over the years, many-world interpretations have been refined. Nowadays, some would argue that the splitting happens only locally and then expands progressively \cite{waegell2020}. In a sense, it has come closer to a relational interpretation.

Still, in my opinion, the many-world interpretations commit the same kind of mistake that was committed by Ptolemy, namely, it attributes to the world some properties which are only perspectival: Ptolemy took epicycles as the real motion of planets, whereas it is nothing but his own motion projected in the sky. Similarly, in many-worlds, one thinks that the world splits in branches, whereas it may only indicate that the points of views are many.

At least, this is what the relational interpretation (RQM), as understood by Rovelli in 1995, would claim \cite{rovelli1996b}. The two observers get different final states, and that’s it. The state of a system is not something absolute, it is only a description relative to some observer. So, when an experiment is described, it is crucial to make explicit the reference observer, otherwise paradoxes are encountered. If one sticks to a given observer, no contradiction shows up. Such an interpretation is \textit{epistemic}, as opposed to \textit{ontic}, meaning that it does not grant an objective reality to the wave-function, but understands it rather as a state of knowledge (or belief). Is such an interpretation so surprising?

\begin{figure}[h]
\includegraphics[width = 0.7 \columnwidth]{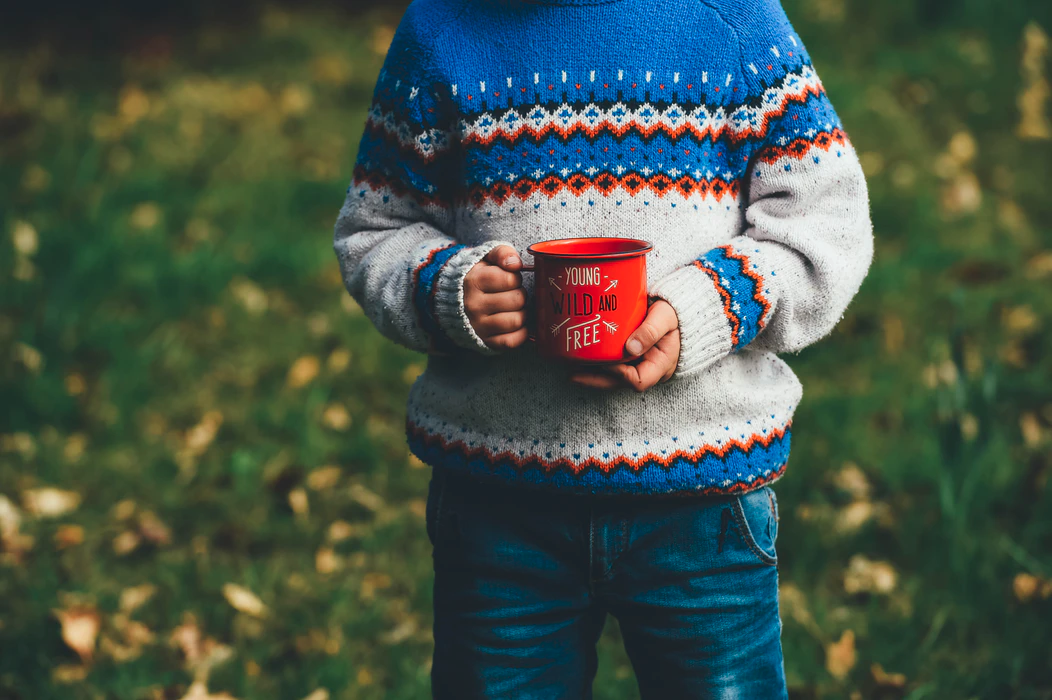}
\includegraphics[width = 0.7 \columnwidth]{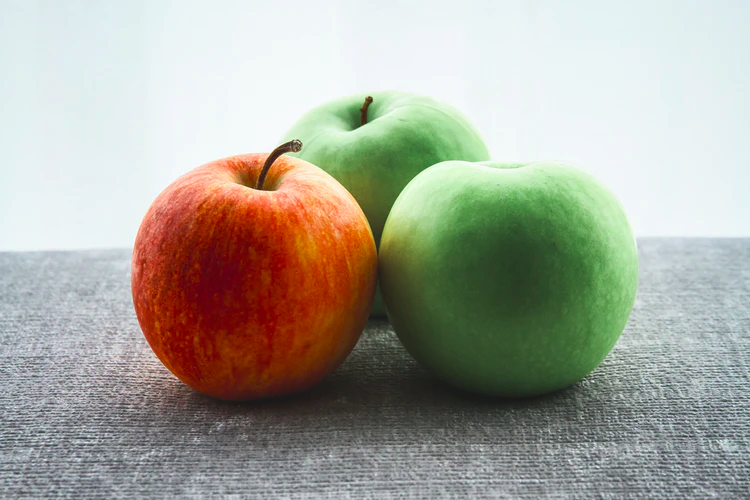}
\caption{External and internal properties. Top: left-handed or right-handed cup? Bottom: colourful apples (figuring quarks). Photos respectively from \cite{spratt} and \cite{wong}}.
\label{fig:apple-cup}
\end{figure}

Maybe not. If we are shown a cup like in figure \ref{fig:apple-cup}, we see a left-handed cup... but the boy holding it sees a right-handed cup. So it is pretty obvious that different observers give a different account of the same system. To the question, "where is the handle?", one answers "left", the other "right". It is just a particular example of kinematic relativity, and it is not surprising at all.
It becomes surprising when relationality is claimed for properties which are "internal", i.e. not referring to space-time localisation, as opposed to "external". It seems strange to say that the colour of a quark depends, not only on the quark, but also on the observer who is looking at it (see figure \ref{fig:apple-cup}). But in RQM, the distinction between internal and external degrees of freedom is only superficial.

The measurement of a property establishes a correlation between the system and the observer, so that the measurement of the colour of a quark can also be regarded, symmetrically, as the measurement \textit{of} the observer \textit{by} the quark. Physics describes relations between systems rather that systems themselves. This is the relational core of RQM. QM is not only about this (it is also about probabilities, discreteness...), but it includes this crucial relational lesson.

This raises a difficulty about the ontology. What is primordial, systems or relations between systems? My opinion is that the notion of system is secondary. A system is an abstract notion useful to recast past measurements whose result is assumed to remain unchanged throughout later experiments. A system is defined by constant properties which serve as a signature to track its identity. For instance, the mass of a particle identifies it as a system to the extent that it remains constant when going through a Stern-Gerlach apparatus. So, the definition of a system relies on a peculiar choice of observables. This choice is not absolute, but rather contingent to a given set of experiments. A particle, identified by its mass, constitutes a proper system as long as it does not interact with an antiparticle, in which case the particle disappears. For such an annihilation process, the notion of system would have to be reconsidered to include the all Fock space.

The primacy of relations upon systems is a general metaphysical statement, not specific to QM or GR, as the following simple example shows. A cup is recognised as an objective system by considering that it is the same whatever the perspective it is looked from. What comes first to experience is the perspectival view on the cup. The idea of a cup is a later reconstruction. Formally, it is tempting to say that the cup \textit{is} the equivalence class of all the perspectives on the cup. This view is a funny way to define a cup, and not completely convincing because no-one has ever poured tea in an equivalence class. There is a gap between the formal definition and the practical use of the term. The same gap makes us sceptical with the definition of space-time as an equivalence class of solutions of the Einstein equations. But formally it is correct. A given physical theory will additionally specify some mathematical structure on top of the equivalence class. For instance, in QM, the equivalence class of a system has the structure of a Hilbert space, which is not the case in the example of the (classical) cup. A canonical example is given by the definition of a particle as an irreducible representation of the Poincaré group. This group encodes all the various perspectives that an observer can have on a particle. Its irreducible representations are characterised by two numbers, the mass and the spin, which label univocally, and thus \textit{define} the particle as an objective system. This is very concretely how the identity of particles is tracked in bubble chambers or particle accelerators.

\section{Conclusion}
\label{sec:conclusion}

We have seen that the relational novelties introduced by the fundamental theories of physics show up in very different manners. A major difference that should be pointed out lies at the formal level. The relationality of GR can be captured in the fact that it is a gauge theory. This is mathematically very precise: there is a gauge group (the diffeomorphisms) acting on a set of fields and the observables are gauge-invariant functions of these fields. Such is not the case in QM. The relational interpretation has not yet been fixed in such a formal shape, although recent attempts to describe changes between quantum reference frames may be seen has important steps in that direction \cite{hamette2020, vanrietvelde2020}. To a large extent, the language, the formalism and the domains of these fundamental theories are thus incomparable. On the way to quantum gravity, we may expect relationality to show up from both QM and GR. But at this stage, it is very hard to say whether one should consider relationality as a guiding principle for the quest of quantum gravity.

Instead, I would like to argue from an epistemological point-of-view, that the various instantiations of relationality pertain to the same movement in the progress of physical theories. In all of the fundamental theories that I described above, some notions that were previously believed to be absolute have become relative. In GR, the notion of distance between two material points has been made relative to the values of the gravitational field in-between. Thus, a gravitational wave changes the distance between two test-masses. This is contrasting with Newton's theory where the background space-time was thought to be independent of the gravitational force that lived within. Similarly, in QM, the notion of internal properties has been made dependent on the perspective of an observer, whereas a strict independence was previously assumed. These theories have thus revealed a \textit{context} that was implicit before.

This process of "relativisation" of physical concepts is pretty common in physics. For instance, Galileo found that "velocity" is not absolute, but depends on the choice of an inertial frame. Also, in special relativity, the notion of "simultaneity" has been made dependent on the velocity of the reference observer. At an even more basic level, the value attributed to a measurable quantity, like the mass, is not an absolute number, but it depends on the choice of a unit of measure, like the kilogram.

From a linguistic perspective, regarding a theory as a web of allowed words and sentences, the new theories are unravelling unexpected conditionalities in the use of terms like "velocity" or "state". If these conditionalities are ignored, the experiments cannot be described consistently. In the new theories, the same words are still in use, like "simultaneity" or "trajectory", but with new degrees of meaning, which, like back doors, are made to avoid contradiction.

The central role of relationality in the development of physical theories is sometimes overlooked. One often hears that physics consists in finding the "invariants" of nature, in a slow process of objectivisation of reality. This is certainly true, but it is only half of the story. In this paper, we have seen the centrality of the opposite process that consists in finding new "dependencies", uncovering the subjective side of reality.

\begin{acknowledgments}
I thank the organisers of the QISS conference in Hong Kong in January 2020 for giving me the opportunity to deepen and present the ideas contained in this paper.

This publication was made possible through the support of the ID\# 61466 grant from the John Templeton Foundation, as part of the “Quantum Information Structure of Spacetime (QISS)” project (\hyperlink{http://www.qiss.fr}{qiss.fr}). The opinions expressed in this publication are those of the authors and do not necessarily reflect the views of the John Templeton Foundation.
\end{acknowledgments}

\vfill

\bibliographystyle{scipost}
\bibliography{relation.bib}

\end{document}